\documentclass[12pt,twoside]{article}   %For printing on two sides of page
\usepackage[super,sort&compress,comma]{natbib}
\usepackage{fancyhdr}		%Gives headers and footers defined below
\newcommand{\captionv}[3]{\begin{center}\parbox{\textwidth}{\caption[#2]{{\sf #3}}}
        \end{center}}
\usepackage[section]{placeins}   %
\usepackage{graphicx}

\makeatletter \renewcommand\@biblabel[1]{$^{#1}$} \makeatother
\newcommand{\cen}[1]{\begin{center} #1 \end{center}}

\usepackage[mathlines]{lineno}
\usepackage{hyperref}
\hypersetup{ colorlinks,
    citecolor=blue,
    filecolor=blue,
    linkcolor=blue,
    urlcolor=blue
}

\usepackage{xcolor}
\definecolor{gray}{rgb}{0.6,0.6,0.6}
\definecolor{red}{rgb}{0.85,0,0}
\definecolor{green}{rgb}{0,0.85,0}
\definecolor{blue}{rgb}{0,0,0.85}
\definecolor{beige}{rgb}{0.92,0.87,0.78}
\usepackage[all]{hypcap}    %causes link to figures to go to figure, not caption
\usepackage[T1]{fontenc} % to solve the problem with E. Stepien with \k in the bibliography

\usepackage{subfig} % WK - added it

\begin{document}

\cen{\sf {\Large {\bfseries Realistic Total-Body J-PET Geometry Optimization - Monte Carlo Study } \\  
\vspace*{10mm}
Jakub Baran$^{1,2,3}$, Wojciech Krzemien$^{4,2,3}$, Szymon Parzych$^{1,2,3}$, Lech Raczyński$^{5}$, Mateusz Bała$^{5}$, Aurélien Coussat$^{1,2,3}$,  Neha Chug$^{1,2,3}$, Eryk Czerwiński$^{1,2,3}$, Catalina Oana Curceanu$^{6}$, Meysam Dadgar$^{1,2,3}$, Kamil Dulski$^{1,2,3}$, Kavya Eliyan$^{1,2,3}$, Jan Gajewski$^{7}$, Aleksander Gajos$^{1,2,3}$, Beatrix C. Hiesmayr$^{8}$, Krzysztof Kacprzak$^{1,2,3}$, Łukasz Kapłon$^{1,2,3}$, Konrad Klimaszewski$^{5}$, Grzegorz Korcyl$^{1,2,3}$, Tomasz Kozik$^{1,2,3}$, Deepak Kumar$^{1,2,3}$, Szymon Niedźwiecki$^{1,2,3}$, Dominik Panek$^{1,2,3}$, Elena Perez del Rio$^{1,2,3}$, Antoni Ruciński$^{7}$, Sushil Sharma$^{1,2,3}$, Shivani$^{1,2,3}$, Roman Y. Shopa$^{5}$, Magdalena Skurzok$^{1,2,3}$, Ewa Stępień$^{1,2,3}$, Faranak Tayefiardebili$^{1,2,3}$, Keyvan Tayefiardebili$^{1,2,3}$, Wojciech Wiślicki$^{5}$, Paweł Moskal$^{1,2,3}$ \\ }

$^1$Faculty of Physics, Astronomy and Applied Computer Science, Jagiellonian University, Kraków, Poland
$^2$Total-Body Jagiellonian-PET Laboratory, Jagiellonian University, Kraków, Poland
$^3$Center for Theranostics, Jagiellonian University, Kraków, Poland
$^4$High Energy Physics Division, National Centre for Nuclear Research, Otwock-Świerk, Poland
$^5$Department of Complex Systems, National Centre for Nuclear Research, Otwock-Świerk, Poland
$^6$INFN, Laboratori Nazionali di Frascati, Frascati, Italy
$^7$Institute of Nuclear Physics Polish Academy of Sciences, Kraków, Poland
$^8$Faculty of Physics, University of Vienna, Vienna, Austria

%\vspace{5mm}\\
%Version typeset \today\\
}

\pagenumbering{roman}
\setcounter{page}{1}
\pagestyle{plain}
Author to whom correspondence should be addressed. email: jakubbaran92@gmail.com, wojciech.krzemien@ncbj.gov.pl \\
% note, probably best not to use a student's e-mail as it won't be valid for
% very long.

\begin{abstract}
\noindent {\bf Background:} Total-Body PET is one of the most promising medical diagnostics modalities, opening new perspectives for personalized medicine, low-dose imaging,  multi-organ dynamic imaging or kinetic modelling. 
The high sensitivity provided by Total-Body technology can be advantageous for novel tomography methods like positronium imaging, demanding the registration of triple coincidences. 
Currently, state-of-the-art PET scanners use inorganic scintillators. However, the high acquisition cost reduces the accessibility of Total-Body PET technology. Several efforts are ongoing to mitigate this problem. Among the alternatives, the Jagiellonian PET (J-PET) technology, based on axially arranged plastic scintillator strips, offers a low-cost alternative solution for Total-Body PET. \\ 
{\bf Purpose:} The work aimed to compare five Total-Body J-PET geometries with plastic scintillators
%and indicate the next-generation J-PET scanner design, 
suitable for multi-organ and positronium tomography as a possible next-generation J-PET scanner design. \\
{\bf Methods:} We present comparative studies of performance characteristics of the cost-effective Total-Body PET scanners using J-PET technology. We investigated \textit{in silico} five Total-Body scanner geometries, varying the number of rings, scanner radii, and other parameters. Monte Carlo simulations of the anthropomorphic XCAT phantom, the extended 2-meter sensitivity line source and positronium sensitivity phantoms were used to assess the performance of the geometries. Two hot spheres were placed in the lungs and in the liver of the XCAT phantom to mimic the pathological changes. We compared the sensitivity profiles and performed quantitative analysis of the reconstructed images by using the quality metrics such as contrast recovery coefficient, background variability and root mean squared error. The studies are complemented by the determination of sensitivity for the positronium lifetime tomography and the relative cost analysis of the studied setups. \\
{\bf Results:}
The analysis of the reconstructed XCAT images reveals the superiority of the seven-ring scanners over the three-ring setups.  
However, the three-ring scanners would be approximately 2-3 times cheaper.
The peak sensitivity values for two-gamma vary from 20 to 34 cps/kBq and are dominated by the differences in geometrical acceptance of the scanners. 
The sensitivity curves for the positronium tomography have a similar shape to the two-gamma sensitivity profiles.
The peak values are lower compared to the two-gamma cases, from about 20-28 times, with a maximum value of 1.66 cps/kBq. This can be contrasted with the
50-cm one-layer J-PET modular scanner used to perform the first in-vivo positronium imaging with sensitivity of 0.06 cps/kBq. \\
{\bf Conclusions:}
The results show the feasibility of multi-organ imaging of all the systems to be considered for the next generation of TB J-PET designs.
Among the scanner parameters, the most important ones are related to the axial field-of-view coverage.
The two-gamma sensitivity and XCAT image reconstruction analyses show the advantage of seven-ring scanners. However, the cost of the scintillator materials and SiPMs is more than two times higher for the longer modalities compared to the three-ring solutions.
Nevertheless, the relative cost for all the scanners is about 10-4 times lower compared to the cost of the uExplorer. 
These properties coupled together with J-PET cost-effectiveness and triggerless acquisition mode enabling three-gamma positronium imaging, make the J-PET technology an attractive solution for broad application in clinics.
%Construction of the Total-Body J-PET scanner is currently in progress.
\end{abstract}
%\note{This is a sample note.}

\newpage     %may or may not be needed

\setlength{\baselineskip}{0.7cm}      %double spacing		

\pagenumbering{arabic}
\setcounter{page}{1}
\pagestyle{fancy}
\section{Introduction}

Positron emission tomography (PET) is a gold standard diagnostic modality enabling metabolic imaging of pathological tissues~\cite{schmall2019,alavi2021unparalleled,alavi2020update}. Presently, the majority of PET machines offer an axial Field-Of-View (FOV) of approximately 20-25~cm  with a single bed position. To perform an image of the entire patient's body, a series of scans acquired with different bed positions is necessary. The new generation of Total-Body (TB) PET scanners~\cite{Cherry2017,spencer2021performance,karp2020,prenosil2022performance} allows for simultaneous imaging of the whole human body, presenting new perspectives in dynamic imaging, kinetic modelling \cite{vandenberghe2020state,vandenberghe2022potential,zhang2020total,wang2022total}, positronium lifetime imaging~\cite{moskal2021positronium,moskalperspectives,moskal2019feasibility}, and simultaneous multi-tracer imaging~\cite{beyene2023exploration}.  

The usage of inorganic L(Y)SO scintillators, while popular, results in high costs of the existing TB scanners, estimated to be in the range of about $\$$10 million or more~\cite{Cherry2018}. The high price reduces the accessibility of TB technology for hospitals and research facilities. To reduce the TB scanner cost~\cite{vandenberghe2022potential}, various approaches have been proposed, including the reduction of the scintillator thickness~\cite{surti2020total,surti2013}, rearrangement of the scintillators to the sparse configurations~\cite{zhang2019sparse,zein2020physical},
use of the BGO crystals combined with the Cherenkov photon signal measurement for timing information
~\cite{cates2019,gundacker2020experimental} or the use of plastic scintillators~\cite{Kowalski2018, moskal2021simulating}. The use of plastic scintillators substantially decreases the price, as they are more than an order of magnitude less expensive than L(Y)SO crystals.~\cite{moskal2020prospects}. 
Moreover, the cylindrical arrangement of the long scintillator strips allows the positioning of the readouts mainly at the ends of the cylindrical rings in contrast to crystal-based PET detectors where coverage of the full cylindrical surface is necessary. In consequence, the amount of 
required silicon photomultipliers (SiPMs), which constitute an important part of the overall scanner price, are greatly reduced.  

%The adoption of plastic scintillators significantly influences the TB system's cost, with the cost of scintillators accounting for approximately 50\% of the overall scanner price in the case of L(Y)SO crystals. The use of plastic scintillators substantially decreases the price, as they are about 80 times less expensive than solid crystals for the 1~cm$^3$~\cite{moskal2020prospects,vandenberghe2020state}. Moreover, the SiPMs and readout constitute another \~30~\% of the TB scanner price. The use of long plastic scintillators allows the positioning of the readouts mainly at the ends of the cylindrical rings in contrast to crystal-based PET detectors where coverage of the full cylindrical surface is necessary.

A cost-effective, portable and modular PET scanner (J-PET) with an extended 50-cm long Axial Field of View (AFOV) is currently in operation at the Jagiellonian University in Kraków.~\cite{tayefi2023evaluation} 
%The scanner is composed of a single layer of 24 independent modules arranged axially. Each module consists of 13  EJ-230 (ELJEN Technology) plastic scintillator strips with the dimensions 24 x 6 x 500~mm$^3$. Each module forms a lightweight, independent unit (about 2 kg) which can be easily transported and assembled in various configurations.  The annihilation photons passing through the plastic scintillator strips interact predominately via Compton scattering. Deposited energy is converted into scintillation light which is then collected at the ends of the scintillators by the SiPMs and read out by fast, customized on-board front-end electronics enabling time-of-flight (TOF) measurement~\cite{raczynski2017calculation}.
The application of the J-PET scanner extends beyond the standard medical two-photon imaging~\cite{moskal2021simulating}.
It provides the capability to conduct multi-gamma tomography studies such as positronium imaging~\cite{moskal2021positronium} and simultaneous multi-tracer imaging~\cite{beyene2023exploration}. Furthermore, it is utilized in fundamental physics studies on quantum entanglement \cite{hiesmayr2019witnessing} and studies of discrete symmetries in nature~\cite{Moskal:2021kxe,moskal2024discrete}. Additionally, it is used in proton beam range monitoring in hadron radiotherapy \cite{baran2019studies,baran2024feasibility,borys2022protheramon,brzezinski2023detection}, and PET data reconstruction methods development~\cite{shopa2021optimisation,raczynski20203d}.

The primary objective of this study is to analyse and compare the performance of five realistic geometry options for designing the new Total-Body (TB) modality based on the J-PET modular technology.
The main functional goal is to develop a cost-effective prototype designed to leverage the TB scanner technology,  enabling e.g. multi-organ imaging. Simultaneously, the sensitivity should allow performing positronium mean lifetime tomography which can deliver complementary information to the currently used standardized uptake value image
~\cite{moskal2020performance,moskal2021positronium,shibuya2020oxygen}.

For further insights into the potential clinical applications of J-PET technology, detailed information can be found in the article~\cite{moskal2020prospects}.

The selection of the particular tested designs including scintillator lengths, gaps between detectors and radii is driven by the constraints imposed by the physical properties of currently available scintillators, the shape of the front-end electronics, and photomultipliers.  Special emphasis is put on the realism of the modelled geometries, with attention to many details abstracted in previous studies. E.g. the previous studies used idealized cylinders with tightly placed scintillators.  
To make the simulation conditions more realistic, we take into account the size of the front-end electronics, and the gaps between adjacent rings, as well as inactive detector material. It has been shown that larger gaps are feasible for 3D PET and can be particularly effective in the design of TB-PET scanners~\cite{yamaya2009multiplex,daube2020performance}.
Additionally, the length of the scintillator strips is restricted to 686.4~mm and 330.0~mm, to improve the time resolution and light yield in the scintillator which is strongly attenuated in longer strips~\cite{Smyrski2017}. In consequence, longer scanners are constructed by combining adjacent rings of cylindrical strips. 

We also introduce the Wavelength Shifter (WLS) layer which improves the precision of the reconstructed position along the scintillator strip~\cite{Smyrski2017}. We developed dedicated GATE modules to handle the details of the new geometry including the WLS.
In contrast to previous studies~\cite{Kowalski2018,moskal2021simulating}, our current work centres on the assessment of the quality of the reconstructed images based on the Monte Carlo simulations of the extended cardiac-torso phantom (XCAT).
Furthermore, we expand the previously used metrics such as contrast recovery coefficient (CRC) and background variability (BV), by adding the Root Mean Square Error (RMSE), and the Q metric which combines the CRC and BV.  Another innovation involves the implementation and application of the bootstrap list-mode method to estimate the metric uncertainty.
Our studies are supplemented by the sensitivity curves for the conventional two-gamma tomography, as well as for the positronium mean lifetime tomography.

%In this study, a different approach has been undertaken. 

%For the purpose of the  
%Beyond sensitivity, %SF and NECR,

%we assess the quality of the reconstructed images simulating the NEMA IEC and extended cardiac-torso  phantoms (XCAT). 

% We start underlining what is new -> especially we present for the first time XCAT studies 
%Second of all we make it such that we really try to do it realistic. Dedicated modules developed for this studies
% this is  supplemented by NEMA IEC phantom and other parameters such as sensitivity etc! 

%Our previous studies focused on the J-PET scanner performance for idealized geometries. The first investigation~\cite{Kowalski2018} considered scanners with axial FOV up to 100~cm. The next studies~\cite{moskal2021simulating} were dedicated to 140~cm and 200~cm long prototypes. In all cases, the scanner model consisted of long plastic strips tightly arranged in a cylinder. In the early article~\cite{Kowalski2018} scanner-based characteristics were determined such as sensitivity, spatial resolution, scatter fraction (SF) and noise equivalent count rate (NECR). For the latter publication~\cite{moskal2021simulating}, the image quality metrics for the NEMA IEC phantom were estimated for the 140~cm long axial FOV scanner. 
%

Based on the performed analysis, the TB J-PET scanner setups are compared.
Currently, works toward the construction of the TB J-PET scanner are ongoing~\cite{moskal2021simulating,Niedzwiecki:2017nka,kaplanoglu2023cross}.

\section{Materials and Methods}
\subsection{Monte Carlo simulation parameters}

J-PET scanner geometries were modelled using GATE v9.0~\cite{sarrut2021advanced,sarrut2022opengate}, based on Geant v4.10.7.1~\cite{agostinelli2003geant4}. An additional layer of Wave Length Shifters (WLS), which improves the estimation of the axial coordinate of the photon interaction point~\cite{Smyrski2017}, was incorporated into the simulation model. The default GATE digitizer code was extended by a dedicated module to allow the simulation of the signals registered in the WLS layers.
In all the simulations the em\_livermore\_polar physics list was used, which is the standard choice for all J-PET-related MC simulations~\cite{moskal2021simulating}. The tracking of optical photons was not included in the simulations to reduce the computation time. In the XCAT and conventional sensitivity simulations, the $\beta^{+}$ source decay is not simulated directly and simulation starts at the emission of the back-to-back photon pairs. 
%The direction of the emission is randomized isotropically. The energy of the initial photons is set to $511$~keV.

\subsection{Phantoms}
We simulated XCAT~\cite{segars20104d}, the extended sensitivity ~\cite{moskal2021simulating} and the positronium sensitivity phantoms. The activity of the male XCAT phantom~\cite{segars20104d} was prepared ~\cite{zincirkeser2007standardized, silva2015simulated} to mimic the $^{18}$F-FDG distribution within the human body. Additionally, two hot spheres (diameter = 1.0 cm) positioned in the lung and in the liver were incorporated in the phantom simulations. The contrast between the hot region and the background activity was set to 16:1 and 3:1 for the lungs and liver, respectively. The overall activity of the phantom was equal to 50 MBq and the acquisition time was set to 600~s.

%In addition, 22~cm long NEMA IEC phantom is built out of four high-activity (denoted as {\it hot}) and two low-activity (denoted as {\it cold}) spheres. The phantom was positioned isocentrically within the scanners. Hot spheres of 10~mm (Sphere 10), 13~mm (Sphere 13), 17~mm (Sphere 17) and 22~mm (Sphere 22) diameters and cold spheres of 28~cm and 37~mm diameters were simulated. The ratio between the hot spheres and the background was set to 4:1. The total activity of 59 MBq was simulated and the acquisition time was set to 500~s.

The extended sensitivity phantom consists of the 2-meter linear source positioned in the centre, along the long axis of the cylindrical scanner. For each simulation, the activity of 10~MBq and measurement time of 1000~s were used. 

For the aforementioned scanners, the back-to-back 511 keV gamma source has been used.

Additionally, the positronium sensitivity phantom consists of a 183 cm-long linear source situated in the transaxial centre of the scanner. In the simulation, the $^{44}$Sc-like source was modelled by using a para-positronium source with enabled deexcitation photon of energy set to 1160 keV.

%In order to calculate sensitivity of investigated geometries a 183 cm long 44Sc – like linear source situated in transaxial center of the scanner was simulated with the GATE software. The simulation was performed assuming a para-positronium source with enabled deexcitation photon of energy set to 1160 keV. The simulated interaction were further plugged into the external coincidence sorter allowing triple coincidence creations. A valid positronium coincidence was defined when two photons originating from the annihilation event were registered within 200 keV and XXX keV, and third photon fell into energy region higher than XXX keV. Such triple registration had to be in coincidence within X ns time window. Additionally, to obtain only true coincidences, conditions such as: originating from same annihilation-deexcitation event or lack of prior interactions of any of three photons were examined. The sensitivity at center was calculated as a ratio of registered true coincidences originating from within central 3 cm region to the activity present within it.

\section{Total-Body J-PET scanner geometries}

%The annihilation photons passing through the plastic scintillator strips interact predominately via Compton scattering. Deposited energy is converted into scintillation light which is then collected at the ends of the scintillators by the SiPMs and read out by fast, customized on-board front-end electronics enabling time-of-flight (TOF) measurement~\cite{raczynski2017calculation}.

Five  TB J-PET scanner configurations were studied. Scanners with varying numbers of rings, the length of the gap between the subsequent rings, scanner radii, and scintillator cross-sections were investigated. The selection of these specific designs, such as scintillator lengths, gaps between detectors, and radius was determined by the constraints imposed by the physical properties of the available scintillators and photomultipliers, dimensions and shape of the front-end electronics, and other materials.                                                             

%The scanner is composed of a single layer of 24 independent modules arranged axially. Each module consists of 13  EJ-230 (ELJEN Technology) plastic scintillator strips with the dimensions 24 x 6 x 500~mm$^3$. Each module forms a lightweight, independent unit (about 2 kg) which can be easily transported and assembled in various configurations.  The annihilation photons passing through the plastic scintillator strips interact predominately via Compton scattering. Deposited energy is converted into scintillation light which is then collected at the ends of the scintillators by the SiPMs and read out by fast, customized on-board front-end electronics enabling time-of-flight (TOF) measurement~\cite{raczynski2017calculation}.

\begin{figure}[htbp]
\centering
\includegraphics[width = 0.45\textwidth]{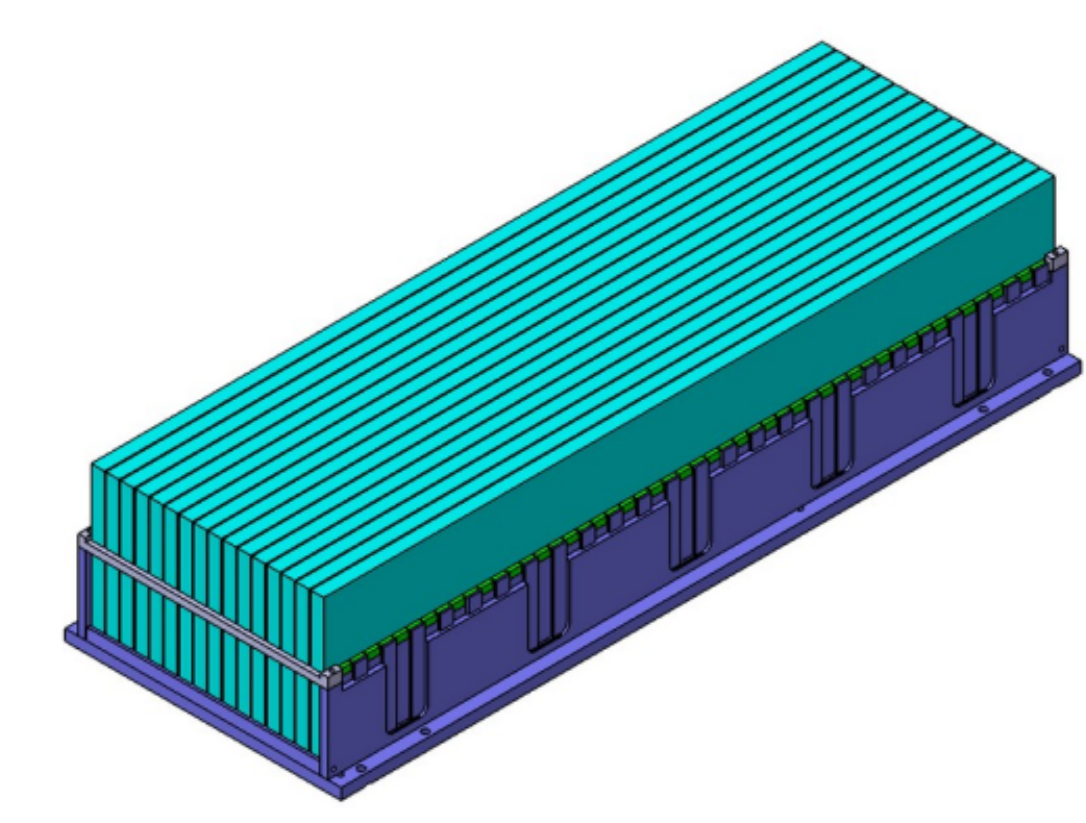}
\includegraphics[width = 0.45\textwidth]{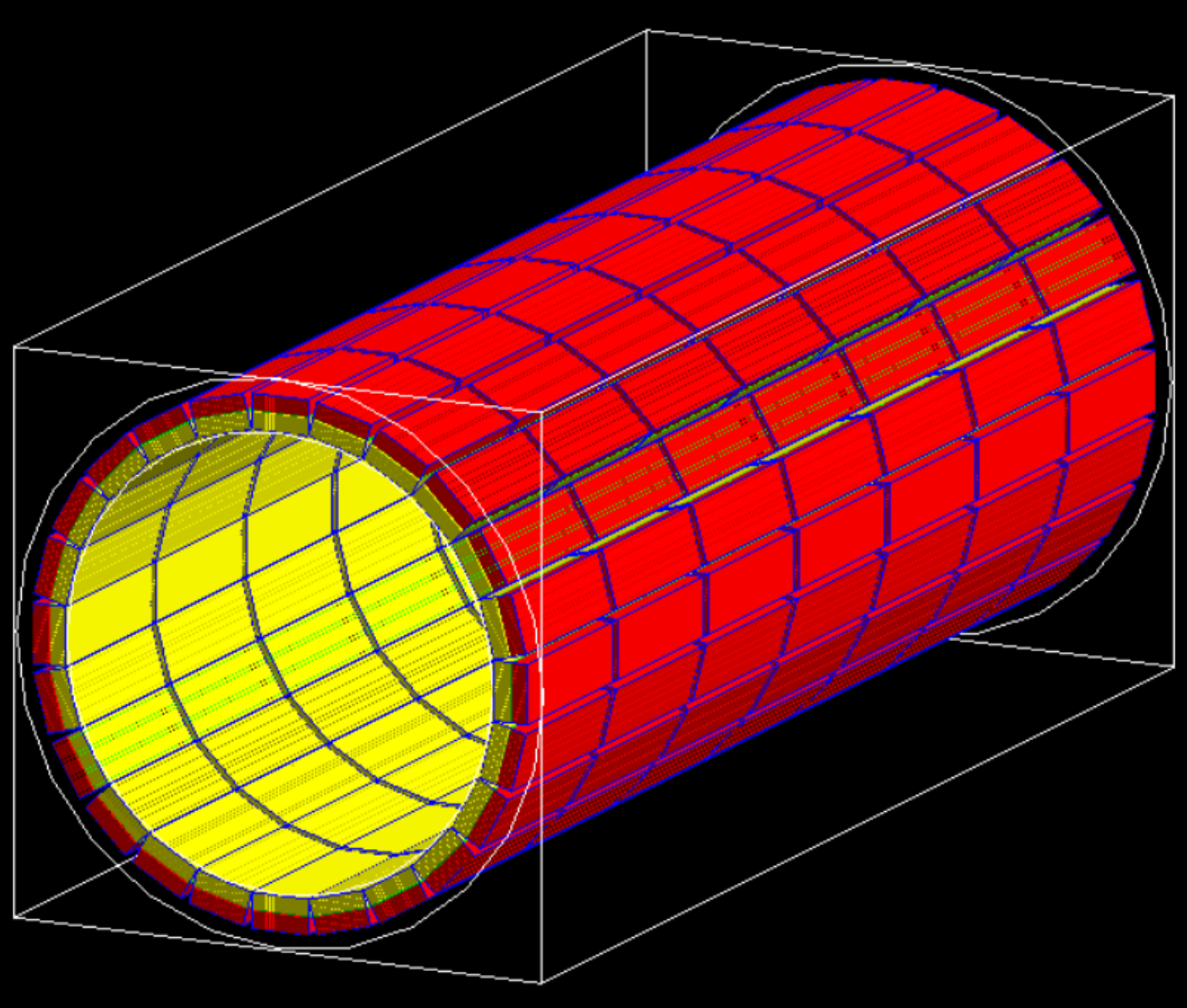}
\captionv{\textwidth}{Visualisation of the scanner geometry.}{(Left) Visualization of the single module, made of plastic scintillators (light blue) with the WLS layer (green elements) and the casing (violet - not included in the simulation model). (Right) Visualisation of the seven-ring S5 scanner with a total FOV of 243 cm. The length of the scintillators in each ring is equal to 33 cm. The gap between adjacent rings is equal to 2 cm. Two layers of the scanners are shown (yellow and red strips).
\label{fig:geometry}
}
\end{figure}

All total-body geometry scanners under consideration are based on the J-PET portable module concept - an independent detection unit composed of plastic scintillator strips (see Fig.~\ref{fig:geometry} left Panel).
The chemical composition of the scintillator corresponds to the commercial EJ230 scintillator, which is used in the existing J-PET scanner prototypes~\cite{moskal2021positronium, moskal2021simulating}. The annihilation photons passing through the plastic scintillator strips interact predominately via Compton scattering. Deposited energy is converted into scintillation light which is then collected at the ends of the scintillators by the SiPMs and read out by fast, customized on-board front-end electronics enabling time-of-flight (TOF) measurement~\cite{raczynski2017calculation}.
%The readout is mounted on both ends of the modules, and 
The modules were placed as closely as possible to the front-end electronics configuration. Each ring consists of 24 modules arranged cylindrically.  The length of the scintillator strips is restricted to 686.4~mm and 330.0~mm, to improve the time resolution and light yield in the scintillator which is strongly attenuated in longer strips~\cite{Smyrski2017}. The dependence of time resolution on strip length was incorporated into the simulation model.
The longer scanners are constructed by combining adjacent rings of cylindrical strips (see Fig.~\ref{fig:geometry} Right Panel).  The plastic scintillators are approximately 7 times less dense than LYSO crystals, and hence less effective for registering annihilation photons. We use two layers to compensate for the lower stopping power of the plastic scintillators.  Adding more layers would increase the cost. 
%Also, the multi-layer setup might lead to a higher fraction of the inter-detector scatter coincidences. 
To enhance the precision of the reconstructed position along the scintillator strip, each module is equipped with a layer of WLSs~\cite{Smyrski2017}.  

In all cases, the modules were placed as close to each other as possible with respect to the electronics configuration. The radius difference between S1-S3 and S4-S5 is determined by the larger module cross-section, which required more space and in consequence significantly greater radius compared to the S3-S5 setups. In both cases, the scintillator cross section is set to fit the size of the SIPM active surface equal to 6$\times$6 mm$^2$.

The selection of these scanner geometries allows for studying the impact of several parameters (see Tab.~\ref{tab1}) on the system performance. Scanners S1 and S2 give insight into the influence of the gap between the rings. The 2 cm gap corresponds to the smallest mechanically possible distance, where all the front-end electronics components are tightly packed, while the 5 cm space allows for some flexibility in the mechanical construction.
The difference between the S1 and S3 modalities consists mainly of the two radius values. The use of a smaller radius permits decreasing the number of readout channels, enhances geometrical angle coverage, and leads to better sensitivity.    
On the other hand, a scanner design with a larger diameter leaves more space between the modules simplifying the mechanical support. The larger diameter, especially in the case of the TB scanners, has the additional advantage of improving the patient's comfort during the scan, by lowering anxiety and reducing claustrophobia, which is one of the psychological burdens the patient must face during medical examination~\cite{total_body_pet, Thatcher558P}.
The S4 and S5 scanners are used to study the effect of the scintillator width. Finally, the S3 and S4 modalities have different cross-sections and use different scintillator lengths resulting in changes in timing resolution\cite{Kowalski2018,moskal2021simulating}.

The price of the considered J-PET scanners was estimated using the methodology inspired by the work~\cite{vandenberghe2020state}. The overall cost of the TB scanner can be decomposed into (a) the cost of the CT system, (b) the cost of scintillator materials, (c) the cost of the front-end electronics boards, and (d) the cost of the SiPMs.
The prices of the electronic boards (c) and CT scanners (a) are assumed to be similar for all TB systems.
The plastic scintillators are about 80 times less expensive than L(Y)SO crystals for the 1~cm$^3$ volume.  In addition,  the cylindrical arrangement of the long scintillator strips allows the SiPMs only at the ends, reducing their number. 
For each scanner setup, we calculate the overall volume of the scintillator materials. Moreover, we include WLS layer volume, assuming that the material per volume is 4 times more expensive than the plastic scintillator price. We also estimate the number of required SiPMs by calculating the area covered by them.
Taking into account (b) and (d) terms, we estimate the price reduction factor for J-PET scanner setups. As a reference point, we take the material budget of the uExplorer  scanner~\cite{spencer2021performance}. Therefore, if a scanner has a price reduction factor equal to 2 it means that it would be two times cheaper than uExplorer in terms of overall SiPMs and scintillator price.

\begin{table}[htbp]
\captionv{\textwidth}{}{Scanners properties.  As a reference point for the price reduction factor, we take the material budget of the uExplorer scanner.
\label{tab1}
}
\begin{center}
\begin{tabular}{|c|c|c|c|c|c|}
\hline
\textbf{}&\multicolumn{5}{|c|}{\textbf{Scanner geometry}} \\
\cline{2-6} 
\textbf{Property} & \textbf{\textit{S1}}& \textbf{\textit{S2}}& \textbf{\textit{S3}}& \textbf{\textit{S4}}& \textbf{\textit{S5}}  \\
\hline
\hline
Radius [mm] & 506 & 506 & 425 & 414.65 & 414.65 \\
Axial FOV [mm] & 2099.2 & 2159.2 & 2099.2 & 2430 & 2430 \\
Scintillator &  &  &  &  &  \\
length [mm] & 686.4 & 686.4 & 686.4 & 330 & 330 \\
Scintillator &  &  &  &  &  \\
cross-section [mm] & 25x5.7 & 25x5.7 & 25x5.7 & 25x6.0 & 30x6.0 \\
No of adjacent &  &  &  &  &  \\
rings & 3 & 3 & 3 & 7 & 7\\
Gap between &  &  &  &  &  \\
adjacent rings [mm] & 20 & 50 & 20 & 20 & 20 \\
Price reduction &  &  &  &  &  \\
factor & 9.6 & 9.6 & 9.6 & 4.3 & 3.6 \\
\hline
\end{tabular}
\end{center}
\end{table}

\subsection{Photon detection, coincidence formation and energy threshold}

The front-end electronic response was modelled by the GATE digitizer which converts photon interaction in the scintillator into deposited energy and detection time.

The temporal and energy resolution of the scanner was described by the phenomenological parameterisation of the experimental resolutions.
The energy resolution dependence is parameterised as a $\frac{\sigma(E)}{E}$ fraction which reflects the experimentally determined relation for the plastic scintillator strips~\cite{Moskal:2014sra}. The simulated photon registration time is smeared, event by event, by replacing the event registration time $t_{r}$ by the value obtained from the normal distribution $N(t_{r},\sigma_{t}$), where $\sigma_{t}$ corresponds to the temporal resolution of the scintillator strip. 
Analogically, the registration position along the scintillator strip (z position) is smeared, event by event, by replacing the registered photon position $z$  by the value obtained from the normal distribution $N(z,\sigma_{z})$, where $\sigma_{z}$ corresponds to the positional uncertainty along the scintillator strip of the scanner. For all the simulations the resolution parameters, i.e., the position along the strip and the time were set to $\sigma_{z}$ equal to 2.12~mm \cite{moskal2021simulating} and $\sigma_{t}$ equal to 100~ps (scanners S1, S2 and S3) or 77~ps (scanners S4 and S5)~\cite{moskal2016time}, respectively. Discrepancies in time smoothing reflect the expected time resolution change due to the size of the scintillator strip length~\cite{moskal2016potential, moskal2018feasibility, moskal2021simulating}.

In the case of the XCAT and extended sensitivity phantoms the coincidence time window of 3~ns was used. In contrast to the inorganic detectors, in plastic scintillators, the photons deposit their energy mainly via Compton scattering. The energy selection threshold $E_{thr}$ of 200 keV was set to reduce the fraction of the background coincidence events ~\cite{Kowalski2018, moskal2021simulating}, for which at least one of them undergoes the scattering in the phantom before being registered in the scanner. Only coincidence pairs with the registration photon energy above the energy threshold are considered. 
This selection criterion corresponds to the optimal selection cut applied in the analysis of the data obtained with the J-PET prototype allowing for the reduction of scattering in the patient and the detectors~\cite{Kowalski2018,moskal2021simulating,Moskal:2014sra}. 

For the positronium imaging sensitivity analysis, simulated interactions were further processed by the external coincidence sorter allowing triple coincidence formation. A valid positronium coincidence was defined as two photons originating from the annihilation registered in the deposited energy range from 200 keV to 350 keV, combined with the third photon registration with the deposited energy above 350 keV. Such triple had to be in coincidence within a 3-ns time window. Additionally, to obtain only true coincidences, conditions such as: originating from the same annihilation-deexcitation event or lack of prior interactions of any of the three photons were applied. 
%The sensitivity at the centre was calculated as a ratio of registered true coincidences originating from within the central 3 cm region to the activity present within it.

\subsection{Data preselection and preparation}

For the image reconstruction analysis, only true coincidences were taken into account. 
%In addition, for all the simulations, the numbers of true, scatter and random coincidences were determined.

The background and contrast 3D region-of-interests (ROIs) were used for the image quality metrics determination. ROIs for the quantitative analysis did not include the whole hot region but were morphologically eroded to overcome the partial volume effects. 15 background ROIs were chosen separately for the liver and lung regions. Both, hot sphere and background ROIs, are composed of 52 voxels. To avoid the partial volume effect, an additional constraint, that the background ROI cannot be neighbouring the region of different activities, was applied~\cite{zhang2017quantitative}. 

\subsection{PET image reconstruction}

The image reconstruction was performed with the CASToR software package~\cite{merlin2018castor}. TOF List Mode - Maximum Likelihood Expectation Maximization algorithm with 150 iterations was used. The reconstruction voxel size was 2.5$\times$2.5$\times$2.5\,mm$^{3}$. TOF resolution kernel was modelled as the Gaussian function. Sensitivity and attenuation corrections were included. The multi-Siddon projector with 10 rays was used. Reconstructed images were smoothed with 3D Gaussian post-filter with $\sigma$ set to 2.5 mm. No additional resolution modelling was used in the reconstruction process.

\subsection{TOF resolution optimization}

The CASToR software provides the possibility to use the shift-invariant TOF kernel only. This approach can be sub-optimal, especially while dealing with large FOV scanners, where the kernel shape can change significantly in the axial direction~\cite{shopa2023tof,daube2006influence}.
However, the aim of these studies is the relative performance comparison between scanner configurations, and the usage of the shift-invariant TOF kernel will affect all the investigated setups in a rather equal manner. 

In the J-PET scanners, the TOF resolution is affected by the additional hit registration uncertainty along the axial direction of the strip that effectively smears the overall TOF resolution (see Fig.~\ref{tof_reso}). More details can be found in the supplementary materials.  
\begin{figure}[htbp]
\centerline{\includegraphics[width = 0.49\textwidth]{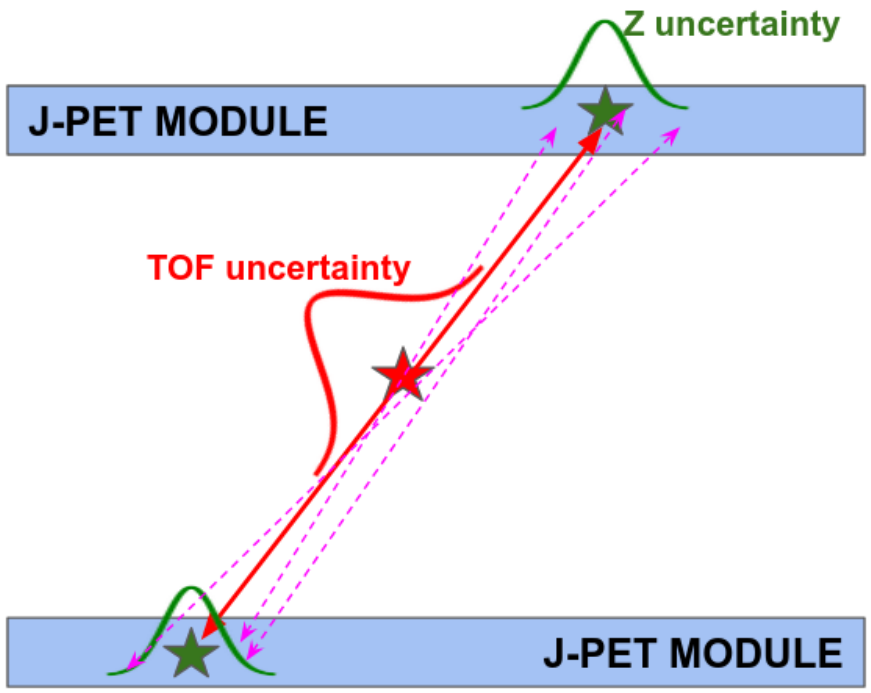}}
\caption{TOF uncertainty sources in the J-PET scanner. Apart from the uncertainty along the line of response (marked in red), additional distortion due to the hit registration resolution along the plastic scintillator (green) is present. Exemplary lines of responses are shown in violet.}
\label{tof_reso}
\end{figure}
As a consequence, it is expected that the optimal width of the shift-invariant Gaussian kernel will be broader than the one determined based solely on the scintillator time resolution properties.
Therefore in the first step, we performed an investigation to select an optimal shift-invariant Gaussian kernel width. The studies were repeated for all setups, to ensure that the common kernel can be chosen.
The selected kernel parameters were applied in all subsequent image reconstruction studies.

%Five different Full-Width Half Maximum (FWHM) kernel values: 115~ps, 230~ps, 365~ps, 500~ps and 750~ps were selected. 
%The investigation was performed with XCAT phantoms.

\subsection{System sensitivity and image quality metrics}

The performance of the scanners has been evaluated based on several criteria including the sensitivity profiles and image quality metrics.

The sensitivity of the scanner S is defined as the rate of detected true coincidences (counts per~s) per unit of the radioactivity concentration. The Monte-Carlo determined ($S_{MC}$) sensitivity in the peak, was compared with the calculated theoretical ($S_{theor}$) value. For that, the geometrical acceptance, detection efficiency of the photon pair, the fraction of events accepted after applying the energy window and the factor accounts for the holes and inactive detector components were considered. In the Supplementary Materials extended explanation of the theoretical sensitivity calculation is given.

The evaluation of the reconstruction performance is based on the metrics defined in the NEMA NU 2–2018 norm for image quality assessment\cite{NEMA:2018}. The procedure is to choose two types of ROIs within the reconstructed image and determine their statistical properties. The first ROI corresponds to the expected high activity ({\it hot}) signal region and the second ROI corresponds to low activity background region(s) ROI. 
The {\it CRC} for a given region of interest is defined as:
\begin{equation}
CRC = \frac{C_{S} -C_{B}}{C_{B}} : \frac{a_{S} -a_{B}}{a_{B}},
\end{equation}
where $C_{S}$, $C_{B}$ is the average number of counts determined for signal and background ROI, and $a_{S}$, $a_{B}$ are the signal and background activities, respectively.

The {\it BV} is defined as a standard deviation ($S_{B}$) calculated for the background ROI normalized to the average counts in the background region:
\begin{equation}
BV = \frac{S_{B}}{C_{B}}. 
\end{equation}

We inspected the additional metric Q, which combines the information from both, CRC and BV, and is defined as~\cite{shopa2021optimisation}: 
\begin{equation}
\label{eq:q}
Q = |CRC - 1| + BV. 
\end{equation}
The Q value range is given by: $Q\in [0,\infty )$. For the perfect image reconstruction in terms of CRC and BV, one expects 1 and 0 for CRC and BV, respectively. By definition, Q would also be equal to 0. 

%In addition, the Root Mean Squared Error (RMSE) metric was used.
We defined the RMSE between two images $I_{1}$ and $I_{2}$ is defined as:
\begin{equation}
RMSE[I_1, I_2] = \frac{1}{N}\sqrt{\sum_{k=1}^{N} (I_1[k] -I_2[k])^2}, 
\end{equation}
where N is the total number of voxels, respectively.

In our studies, the activity emission map was used as a ground-truth reference image.
Before the $RMSE$ calculation, the compared reference and reconstructed images were first normalized to the maximum. Then, the median intensities were calculated for both. Background activity and whole phantom areas were used to calculate medians. Subsequently, the ratio between medians was calculated and finally, the original image was scaled by the ratio.

\subsection{Metric uncertainty estimation}

We use the bootstrap resampling method~\cite{efron1992bootstrap,efron_introduction_1993} to assess the uncertainty of the calculated metrics. Our implementation follows the list-mode-based nonparametric approach proposed in the PET context by the work~\cite{dahlbom2001estimation} and further discussed in the articles~\cite{lartizien_comparison_2010,markiewicz2015assessment}, in which a new sample is formed by 
randomly choosing events with replacement from the original coincidence list obtained from the MC simulations. For each distinctive scenario defined by phantom, scanner setup and parameter settings, 20 bootstrapped samples were prepared and reconstructed. The target metrics values together with their uncertainties are estimated by averaging over the set of bootstrapped realizations.

\section{Results}

\subsection{Sensitivity for two-gamma tomography}
\begin{figure}[htbp]
\centering
\includegraphics[width = 0.65\textwidth]{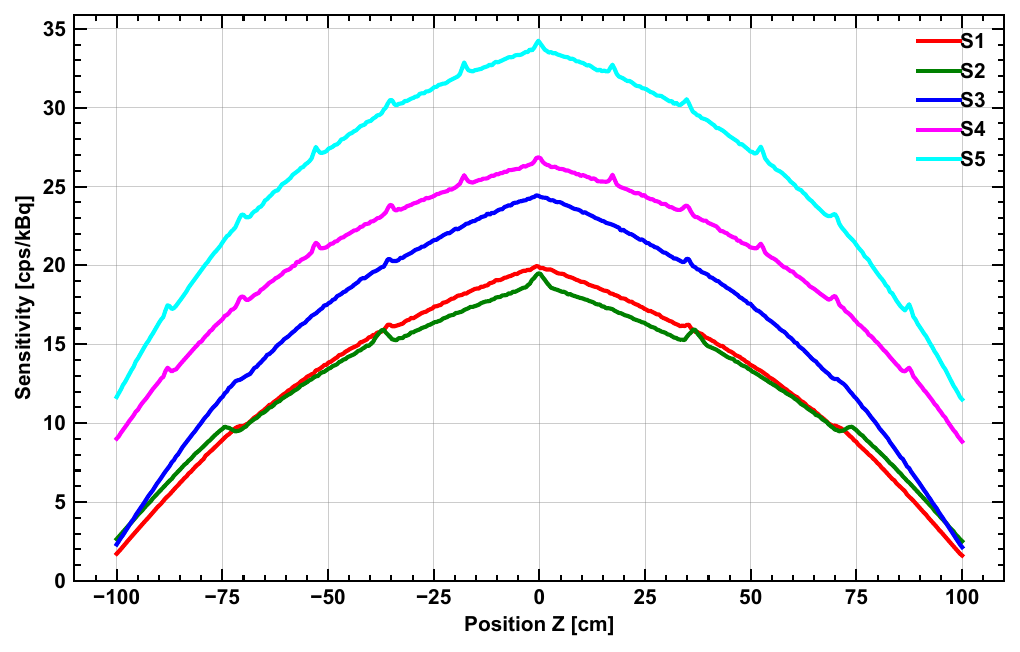}
\captionv{\textwidth}{Sensitivity profiles.}{Sensitivity profiles for tested TB J-PET geometries.
\label{fig2}
}
\end{figure}

Sensitivity plots for all simulated scanners are shown in Fig.~\ref{fig2}. 
The characteristic "spikes" visible in the sensitivity profiles correspond to the centres of the strips and the gaps between adjacent rings.
The sensitivity values in the peak are compared with the analytical calculations in Table~\ref{table_sensitivity}. The Monte Carlo-based results are in reasonable agreement with the analytical calculations. The slight underestimation of the analytical calculations compared to the Monte Carlo ones can be explained by the estimation method of the packing factor, which accounts for holes and inactive detector components without taking into account the depth of the crystal in the estimation.  A detailed explanation of the analytical estimation method can be found in the Supplementary Materials. 
Overall, the greatest sensitivity has been found for the scanner S5 with the maximum at the level of 34 cps/kBq. 
The lowest values are observed for the scanners S1 and S2 with the maximum at the level of 17 cps/kBq.
%(13 cps/kBq). 
The observed difference is a simple consequence of the higher geometrical coverage of the S5 scanner due to its smaller radius and larger AFOV.

\begin{table}
\captionv{\textwidth}{}{Theoretical ($S_{theor}$) and Monte Carlo-calculated ($S_{MC}$) sensitivity in the peak of the sensitivity profile for the source positioned in the scanner centre.
\label{table_sensitivity} 
}
\centering
\begin{tabular}{ |c|c|c| } 
\hline
Type & $S_{theor}$[cps/kBq] & $S_{MC}$[cps/kBq]\\ 
 \hline
          \hline
S1 & 17.61 & 19.95   \\ 
S2 & 17.21 & 19.50   \\ 
S3 & 22.69 & 24.44   \\ 
S4 & 23.21 & 26.85   \\ 
S5 & 30.57 & 34.24   \\ 
 \hline
\end{tabular}
\end{table}

\subsection{Sensitivity for the positronium tomography}

\begin{figure}[htbp]
\centering
\includegraphics[width = 0.65\textwidth]{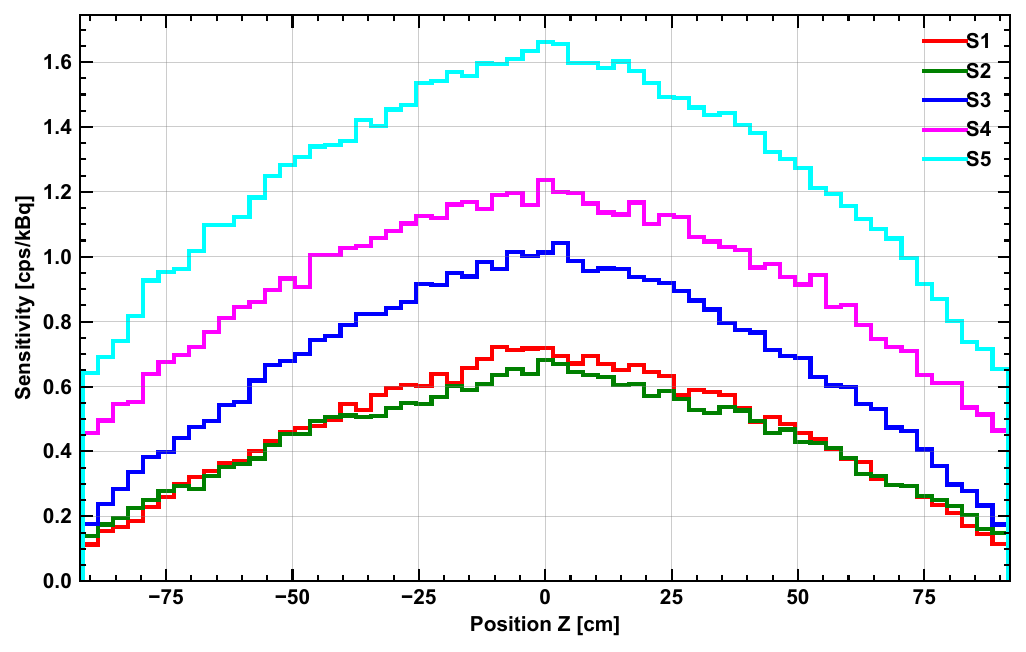}
\captionv{\textwidth}{Sensitivity profiles.}{Multi-gamma imaging sensitivity profiles for tested TB J-PET geometries.
\label{fig2_multi}
}
\end{figure}

The Monte Carlo-based sensitivity plots and sensitivity in peak values for all simulated scanners are shown in Fig.~\ref{fig2_multi} and Table ~\ref{table_sensitivity2}, respectively. 

The profiles are characterized by the same trend as for the two-gamma sensitivity. The greatest values are observed for the S5 scanner (1.66 cps/kBq in peak). On the other hand, the lowest values are observed for the S2 scanner with the maximum at the level of 0.68 cps/kBq. The sensitivity (in the peak) for positronium tomography is about 20-30 times lower than the sensitivity for the two-gamma imaging.

\begin{table}
\captionv{\textwidth}{}{Monte Carlo-calculated ($S_{MC_{2+1}}$) sensitivity in the peak of the sensitivity profile for the source 2+1 positioned in the scanner centre.
\label{table_sensitivity2} 
}
\centering
\begin{tabular}{ |c|c| } 
\hline
Type & $S_{MC_{2+1}}$[cps/kBq]\\ 
 \hline
          \hline
S1 & 0.72  \\ 
S2 & 0.68  \\ 
S3 & 1.01  \\ 
S4 & 1.24  \\ 
S5 & 1.66  \\ 
 \hline
\end{tabular}
\end{table}

\subsection{Number of coincidences}

The numbers of coincidences registered for two-gamma tomography for each scanner and XCAT phantoms are of the order of $10^{7}$ - $10^{8}$ and are given in Table~\ref{table_coincidences}. The greatest number is found for the S5 scanner and the smallest for the S1 scanner. It is an effect of the increased geometrical acceptance and detector efficiency for the thicker (S5 scanner) scintillator layer. 
%The SF and NECR values are given in Table~\ref{table_scatter_fraction_necr}. It is found that the S1 and S2 scanners have slightly lower fraction of scattered coincidences compared to the rest of the scanners. The greatest NECR values are found for the S5 and the lowest for the S2 scanners. 

\begin{table}
\captionv{\textwidth}{}{Number of true, scatter and random coincidences.
\label{table_coincidences}
}
\centering
\begin{tabular}{ |c|c|c|c| } 
\hline
 & \multicolumn{3}{|c|}{XCAT [10$^{7}$]} \\ 
\cline{2-4}
Type & True & Scatter & Random \\
\hline
\hline
S1 & 5.7 & 3.9 & 6.0 \\ 
S2 & 5.5 & 3.7 & 5.8 \\ 
S3 & 8.2 & 6.1 & 9.1 \\ 
S4 & 9.7 & 7.2 & 10.6 \\ 
S5 & 12.6 & 9.5 & 13.5 \\ 
 \hline
\end{tabular}
\end{table}

%\begin{table}
%\captionv{\textwidth}{}{Number of true, scatter and random coincidences for NEMA IEC and XCAT phantoms.
%\label{table_coincidences}
%}
%\centering
%\begin{tabular}{ |c|c|c|c|c|c|c| } 
%\hline
% & \multicolumn{3}{|c|}{NEMA IEC [10$^{7}$]} & \multicolumn{3}{|c|}{XCAT [10$^{7}$]} \\ 
%\cline{2-7}
%Type & True & Scatter & Random & True & Scatter & Random \\
%\hline
%\hline
%S1 & 7.8 & 4.8 & 8.1 & 5.7 & 3.9 & 6.0 \\ 
%S2 & 7.4 & 4.6 & 7.7 & 5.5 & 3.7 & 5.8 \\ 
%S3 & 11.8 & 7.5 & 12.2 & 8.2 & 6.1 & 9.1 \\ 
%S4 & 13.6 & 8.7 & 13.9 & 9.7 & 7.2 & 10.6 \\ 
%S5 & 17.7 & 11.2 & 17.6 & 12.6 & 9.5 & 13.5 \\ 
% \hline
%\end{tabular}
%\end{table}

\subsection{Choice of the optimal TOF kernel}
\begin{figure*}[htbp]
\centering
\includegraphics[width = 0.9\textwidth]{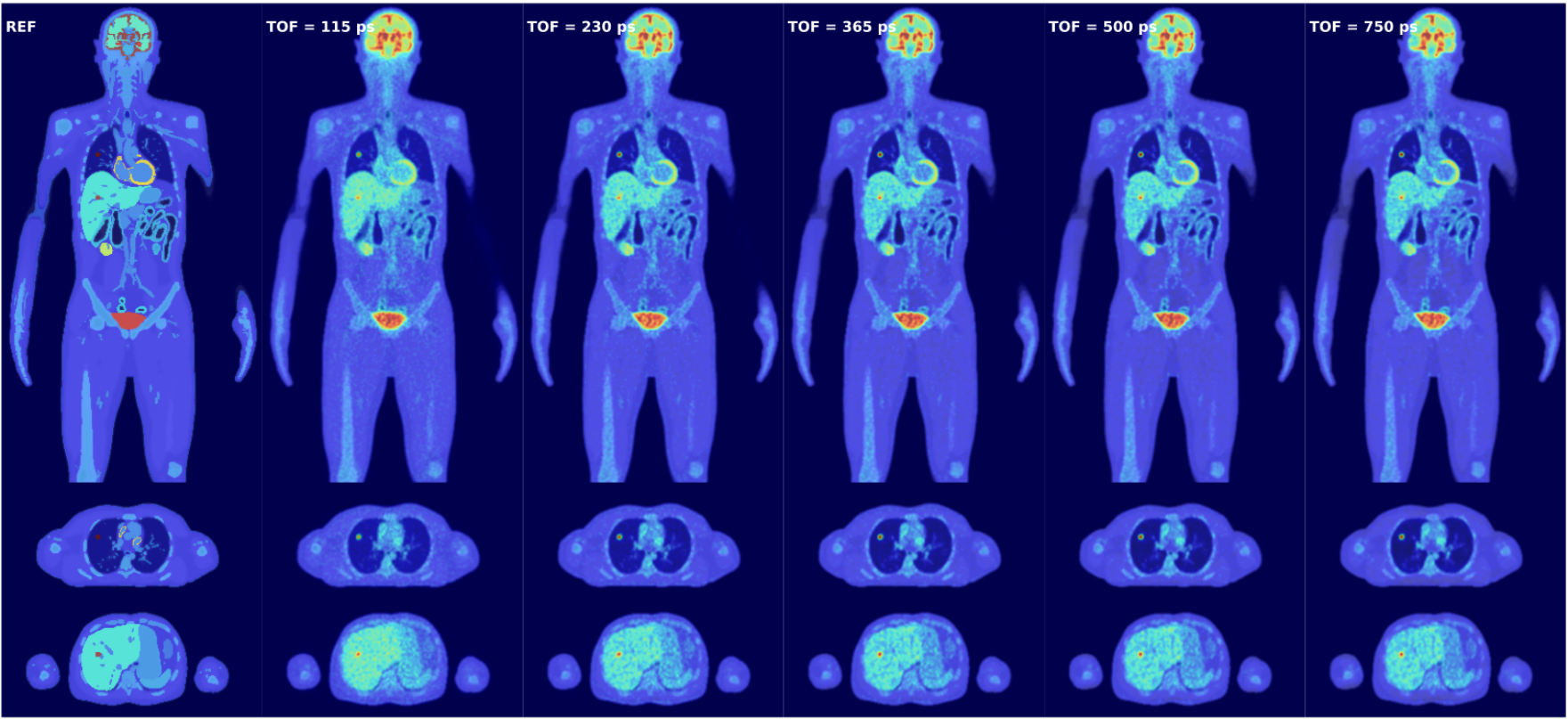}
\captionv{\textwidth}{Reconstructed XCAT images for five different Gaussian TOF widths for S5 scanner.}{Simulated (REF label) and reconstructed XCAT phantom images (TOF label) for five different  Gaussian TOF kernel widths for the sagittal (top panel) and axial (center and bottom panel) views. The center and bottom panels show the slice through the hot spot in the lungs and liver, respectively. PET images are overlayed onto CT scans. Given slices are for S5. 30\textsuperscript{th} iteration images are shown. 
\label{recons_xcat_tofs}
}
\end{figure*}

\begin{figure*}[htbp]
\centering
\includegraphics[width = 0.9\textwidth]{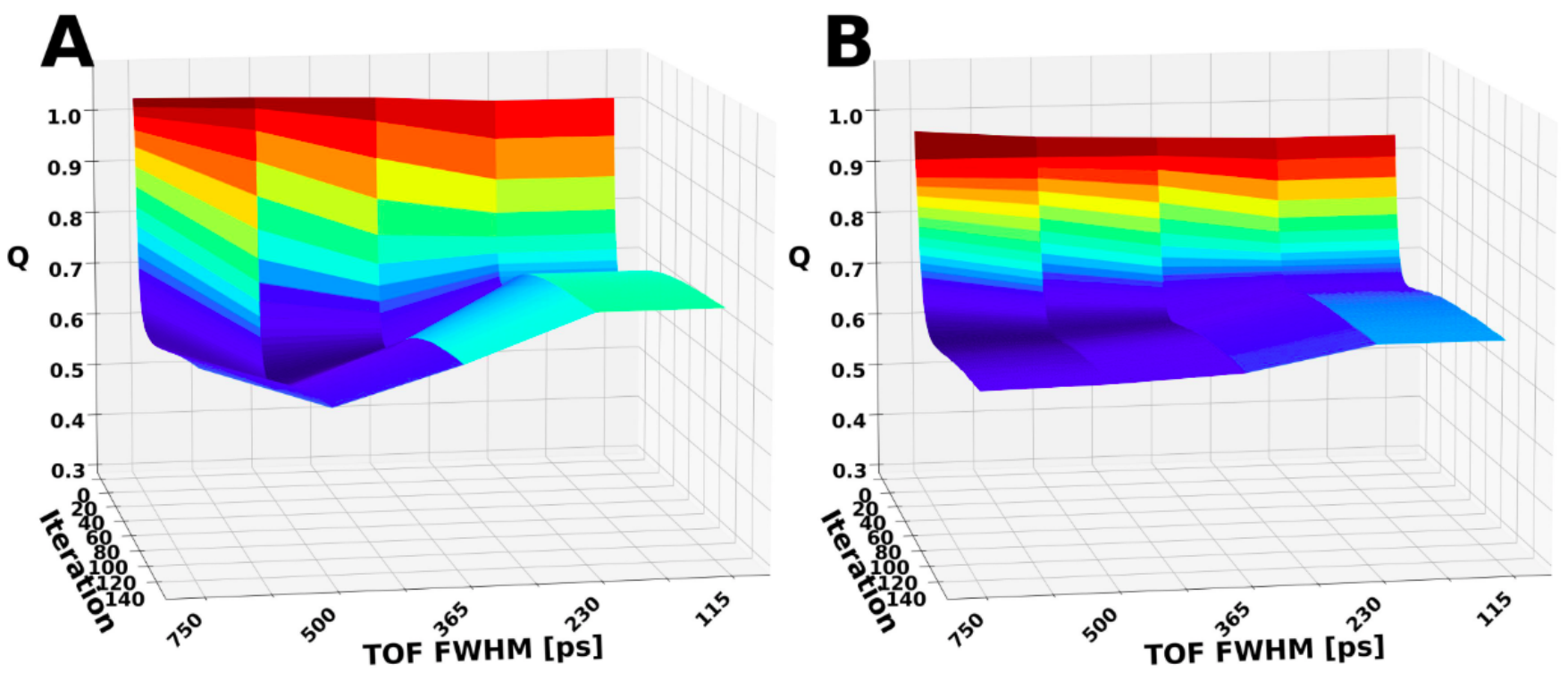}
\captionv{\textwidth}{The Q metric for the S5 scanner.}{The Q metric distributions for lungs (panel A) and liver (panel B) for the scanner S5. 
\label{tof_3D_s5}
}
\end{figure*}

The nominal TOF resolution estimated based on the time-smearing parameters set in the simulations with the scanner S5 corresponds to FWHM of about 166~ps (see Supplementary Materials).
The comparison of the images reconstructed with various Gaussian kernel widths is presented in Fig.~\ref{recons_xcat_tofs}.  
%For all cases, the nominal TOF the simulation, the value of about $\sigma$=77~ps was used.
The visual inspection reveals that the reconstruction with smaller kernel widths (TOF~=~115 ps and TOF~=~230 ps) generates images with higher background and noise levels. Furthermore, the contrast between hot regions and the background is clearer for the images reconstructed with a broader TOF width.
%\begin{figure*}[htbp]
%\centering
%\includegraphics[width = 0.9\textwidth]{Fig4.pdf}
%\captionv{\textwidth}{Reconstructed NEMA IEC images for five different TOF resolutions for S5 scanner.}{Simulated (REF label) and reconstructed NEMA IEC images (TOF label) for five different TOF resolutions for the axial (top panel) and sagittal (bottom panel) view. Given slices are for S5. The 30\textsuperscript{th} iteration images are shown. The TOF values indicated in the figures correspond to the values assumed in the reconstruction. In the simulation the value of about $\sigma$=77~ps was used.
%\label{recons_nema_iec_tofs}
%}
%\end{figure*}
These conclusions are confirmed by analysing
the Q distribution as a function of iteration and applied TOF kernel width presented in Fig.~\ref{tof_3D_s5}. The Q is determined for hot spots. It is found that in all cases the best Q values are observed for the TOF kernel width broader than 230 ps. For five out of six cases (excluding hot spot positioned in the liver) the lowest (the best) Q are found for the TOF~=~365~ps and TOF~=~500~ps. The TOF~=~115~ps obtained the highest (worse) Q characteristic among all the presented cases. It is also noticeable that for all the images for 50\textsuperscript{th} iteration the Q metric reaches the plateau. The same trend was observed for other scanners. 
Corresponding plots are included in the Supplementary Materials.

Taking into account both, the visual inspection and quantitative results, the range of the TOF kernel width between 365~ps to 750~ps gives the best results.
The shapes of the metric plots are very similar and they are preserved for all the scanners.
In the further image quality analysis, we proceed with the images reconstructed with the common Gaussian kernel width set to 500~ps.
%Detailed analysis of the TOF kernel width selection, together with the full set of plots can be found in the appendix~\ref{append1:tof}.

\subsection{Image Quality}

\begin{figure*}[htbp]
\centering
\includegraphics[width = 0.9\textwidth]{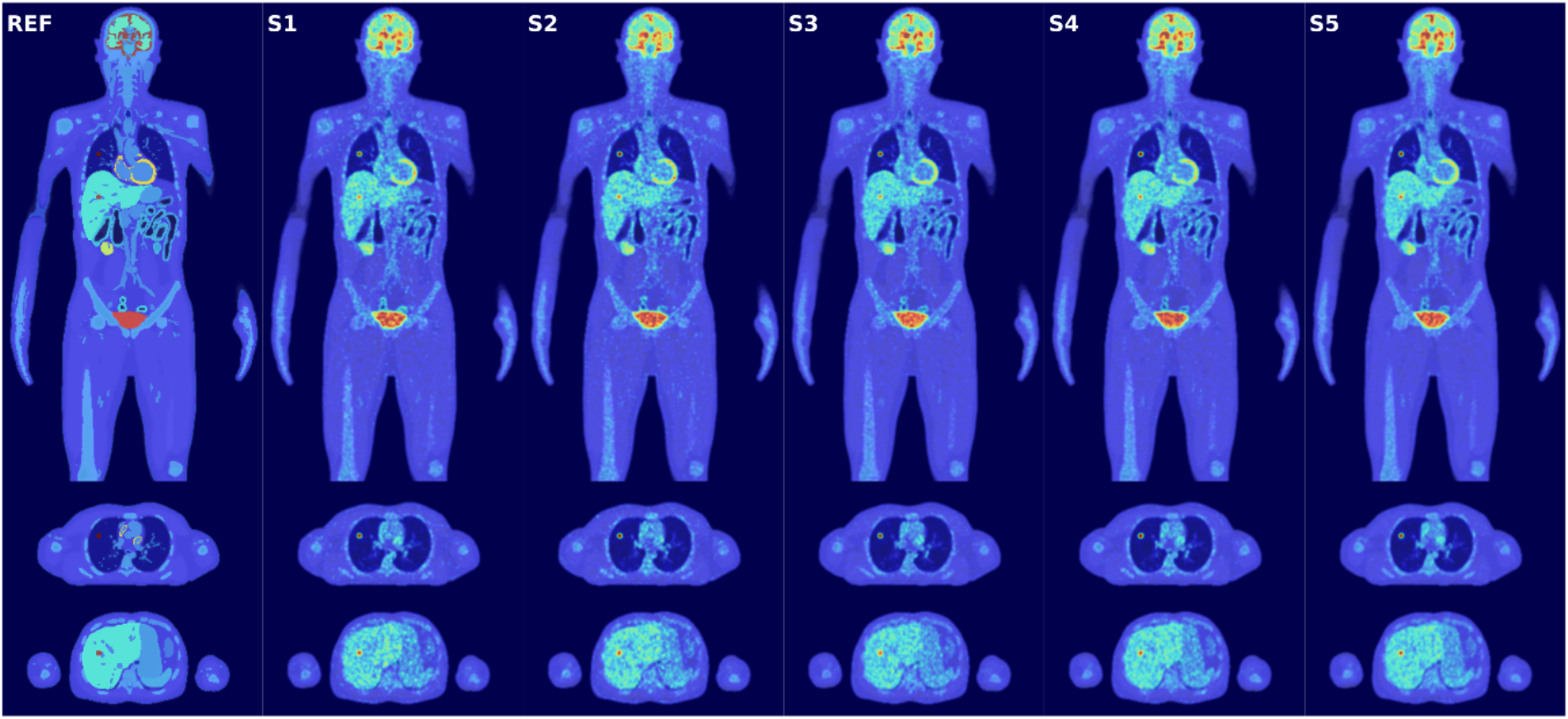}
\captionv{\textwidth}{Reconstructed XCAT images for five different scanners with Gaussian TOF kernel width equal to 500 ps.}{Simulated (REF panel) and reconstructed (S1-S5) XCAT phantom images for five different scanner types for the sagittal (top panel) and axial (center and bottom panel) views. The center and bottom panels show the slices through the hot spot in the lungs and liver, respectively. PET images are overlayed onto CT scans. For each scanner, the 30\textsuperscript{th} iteration image is shown.
\label{recons_xcat}
}
\end{figure*}

Fig. \ref{recons_xcat} presents the images for all the scanners after 30 iterations. The visual inspection indicates that the background noise is higher for the S1, S2 and S3 scanners. It is particularly visible in the shoulder girdles and in the axial slices with hot spots in the lungs and the liver.   At the same time, the contrast values are similar for all investigated scanners.

\begin{figure*}[htbp]
\centering
\includegraphics[width = 1.0\textwidth]{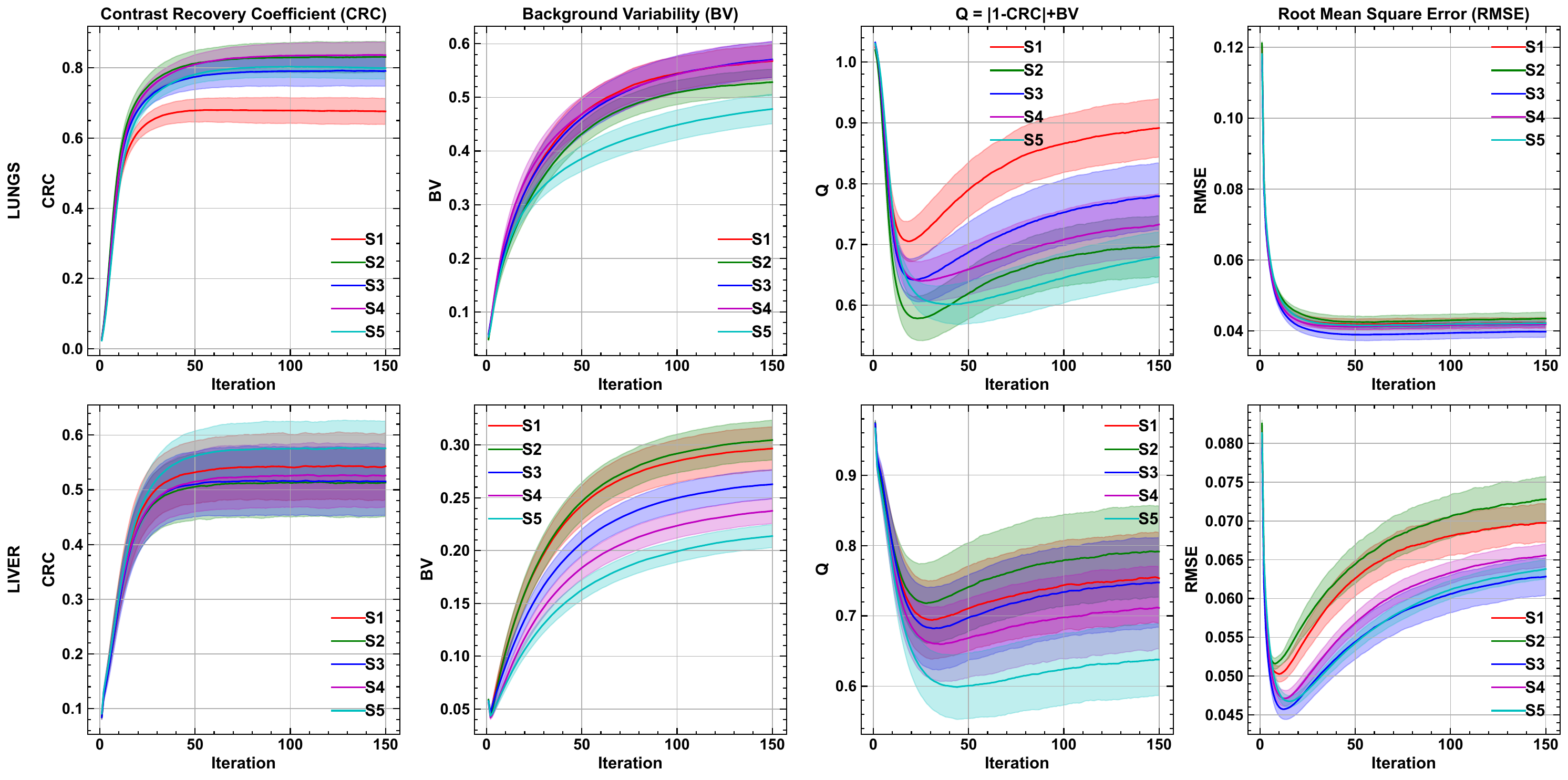}
\captionv{\textwidth}{CRC, BV, Q and RMSE metrics for XCAT for different scanners.}{CRC (first column), BV (second column), Q (third column) and RMSE (fourth column) characteristics for the liver (top row) and lungs (bottom row) regions calculated based on the reconstructed XCAT phantom images for all five scanners. Shaded regions indicate the one standard deviation region.
\label{xcat_500}
}
\end{figure*}

A comprehensive analysis of CRC, BV, Q and RMSE metrics is presented in Fig. \ref{xcat_500}. 
The greatest CRC value is observed for the lungs, although the CRC variation for each ROI among the scanners does not exceed two standard deviations. Vast discrepancies are observed for the BV metrics. Here, scanners S4 and S5 show statistically significant superiority for both the liver and the lung regions over the rest of the scanners. At the same time, the worst results are found for the scanners with the greatest radius - S1 and S2. Above mentioned findings are reflected in the Q metric. As in the case of the BV, the S5 scanner is characterized by the lowest Q metric for both - the liver and the lungs. The same trend is observed for the RMSE metric where the scanner S5 shows an advantage over the results of the other scanners. It can be observed in the liver region. The performance of scanners S1 and S2 is significantly worse than the other scanners. The obtained quantitative results are in agreement with the qualitative, visual inspection.

\section{Discussion}

The most important characteristics determined for the studied scanners are gathered in Table~\ref{table_all}.

\begin{table}
\small
\captionv{\textwidth}{}{Price reduction factor compared to the Explorer material budget, sensitivity in the peak for the two-gamma ($S_{MC}$) and the two+one ($S_{MC_{2+1}}$) sources positioned in the scanner centre, {CRC}, {BV}, and {Q} metrics for the $150^{th}$ iteration in lungs and liver areas.
\label{table_all} 
}
\centering
\begin{tabular}{|c|c|c|c|c|c|c|c|c|c|} 
\hline
Type & Price & $S_{MC}$ & $S_{MC_{2+1}}$ & $CRC_{lungs}$ & $CRC_{liver}$ & $BV_{lungs}$ & $BV_{liver}$ & $Q_{lungs}$ & $Q_{liver}$\\ 
 & factor & [cps/kBq] & [cps/kBq] &  &  &  &  &  & \\ 
 \hline
 \hline
S1 & 9.6 & 19.95 & 0.72 & 0.68(4) & 0.54(6) & 0.57(3) & 0.30(2) & 0.89(5) & 0.75(6)\\ 
S2 & 9.6 & 19.50 & 0.68 & 0.83(4) & 0.51(6) & 0.53(2) & 0.30(2) & 0.70(5) & 0.79(6)\\ 
S3 & 9.6 & 24.44 & 1.01 & 0.79(4) & 0.52(6) & 0.57(3) & 0.26(1) & 0.78(5) & 0.75(6)\\ 
S4 & 4.3 & 26.85 & 1.24 & 0.84(4) & 0.53(6) & 0.57(3) & 0.24(1) & 0.73(5) & 0.71(6)\\ 
S5 & 3.6 & 34.24 & 1.66 & 0.80(3) & 0.58(5) & 0.48(3) & 0.21(1) & 0.68(4) & 0.64(5)\\ 
 \hline
\end{tabular}
\end{table}

In the presented study two-gamma sensitivity profiles for five total-body J-PET scanner geometries (Table~\ref{tab1} and \ref{table_all}) were determined. 
The peak sensitivity values vary from 20 to 34 cps per kBq and are dominated by the differences in the geometrical acceptance of the scanners. 
The maximum peak sensitivity of 34 cps per kBq was found for the scanner S5. 
This value is slightly lower than the result reported in our previous study~\cite{moskal2021simulating}, where for the 200 cm ideal scanner the sensitivity in peak is equal to 38 cps per kBq. 
The difference can be explained by the larger radius of the investigated setup (41.4 cm versus 39 cm) and the inclusion of the gaps between rings and adjacent plastic scintillators. 
Contrasted with the values reported for the state-of-the-art TB scanners: uExplorer (191.5 cps per kBq) and PennPET (55 cps per kBq)~\cite{vandenberghe2020state}, the J-PET scanner sensitivity is lower, however, it can be seen as a significant improvement concerning the current 16–26 cm long PET systems~\cite{moskal2020performance}. 

The sensitivity curves for the positronium tomography have a similar shape to the two-gamma sensitivity profiles.
Since they are calculated based on the triple coincidences formed by the two annihilation and one deexcitation photons, the values in the peak are much lower compared to the two-gamma cases, from about 28 times lower for the S1 and S2 scanners to about 20 times lower for the S5 scanner. 
One can observe, that the sensitivity for triple coincidences rises much faster as a function of AFOV than the corresponding dependence for the two-gamma case. The sensitivity in the peak for positronium imaging rises by 130\% comparing the S5 to S1 scanner. On the other hand, the two-gamma sensitivity in the peak gain raises by only 70\% from S1 to S5.

The highest peak sensitivity for positronium imaging is estimated for the S5 scanner and is equal to 1.66 cps/kBq. This value can be contrasted with the
estimated sensitivity of the positronium imaging in the peak of 0.06 cps/kBq of the J-PET modular scanner, which has been used to reconstruct the first positronium mean lifetime image of the human brain in vivo~\cite{moskal2024_ScienceSubmitted}.
This makes the J-PET TB modality a well-suited scanner for multi-photon imaging~\cite{moskalperspectives,moskal2020prospects}.
%Positronium sensitivity ...
%The sensitivity (in the peak) for positronium tomography is about 20-30 times lower than the sensitivity for the two-gamma imaging.
%If we focus on the 2$\gamma$ + $\gamma_{prompt}$ coincidences necessary for the positronium imaging technique~\cite{moskal2020performance,moskal2021positronium,shibuya2020oxygen}, 
%
%The sensitivity of the J-PET scanner can be further increased by adding the third, outer layer of scintillators, which considering the low cost of the plastic scintillators, makes it an attractive solution. 

The quantitative analysis of the reconstructed image quality was performed based on simulations of the XCAT phantom.
The analysis of the reconstructed images confirms the feasibility of multi-organ imaging with  J-PET TB technology.
%Rather it shows the feasibility of multi-organ imaging.
The visual inspection reveals the superiority of the seven-ring scanners (S4 and S5) over the three-ring setups (S1, S2 and S3) (Fig.~\ref{recons_xcat}). In particular, the noise level is smaller for the scanners S4 and S5. This is confirmed by the quantitative results shown in Fig.~\ref{xcat_500} and in Table~\ref{table_all}. The BV value for the S5 is the lowest for both lesions. 
%The CRC values do not give any advantage to a given scanner. 
BV values are of a similar order as the CRC ones. Thus, both BV and CRC will have similar importance in the Q metric results. For this case, the Q value indicates the scanners S4 and S5 are the optimal geometries for a Total-Body J-PET. Additionally, the RMSE characteristics (for the liver lesions specifically) show the scanners S4 and S5 to best mimic the simulated reference image.

In our study, the simulated scenarios use the same overall activity between investigated scanners and the acquisition time for all the scanners to better reproduce the real conditions.
As a consequence, the reconstructed images and calculated metrics differ in the number of registered coincidences. 
It is plausible that the effect of the reduced background variability 
%in the NEMA IEC and XCAT simulation scenarios 
can be fully explained by the higher statistics acquired by the scanners S4 and S5. 
%S4 and S5 for the scanners with the highest statistics (S4 and S5). 

Note that the presented reconstruction and analysis protocols require further optimization for potential clinical application and it is out of the scope of this paper.

\subsection{Limitations} 

In the image reconstruction process, the sensitivity and attenuation corrections were included.
Image quality could be further improved by the addition of other correction factors such as point spread function (PSF) or depth-of-interaction (DOI) modelling. Indeed, the efforts to develop a dedicated J-PET system response matrix are ongoing~\cite{shopa2023tof}.
Also, no selection criteria for the obliqueness of the accepted line-of-responses were applied. As shown in our previous study, it could improve the contrast and background of the reconstructed images~\cite{kopka2020reconstruction}. 
However, the more accurate system matrix modelling or application of the obliqueness selection criteria would improve the overall metric values e.g. contrast, but would not change the relative trends we observed among the scanner setups.

In our study, only the true coincidences were used for the image reconstruction. Further studies must be carried out to develop the scatter and random correction methods for the J-PET-based TB scanners. 

\section{Conclusions}

We performed comparative studies of five realistic, two-layer TB J-PET scanners, based on the Monte Carlo simulations of the XCAT phantom. 
The overall performance was quantified in terms of CRC, BV, Q and RMSE metrics for the two-gamma tomography. In addition, the two-gamma and positronium imaging sensitivity curves were determined.  
The results show the feasibility of multi-organ imaging of all the systems to be considered for the next generation of TB J-PET designs.

Among the scanner parameters, the most important ones are related to the axial FOV coverage. The influence of the time resolution depending on the scintillator length, positional uncertainty due to different material thickness and other parameters have a secondary effect on the reconstruction quality within the considered values range, and are dominated by the parameters related to the overall registration efficiency such as total length of the active material or scanner radius.
%Przerobic to, bo to nie do konca ma sens, trzeba napisac, ze najwazniejsze czynniki to pokrycie geomentryczne czyli FOV (coverage) i radius, bo to decyduje o statystyce natomiast inne czynniki (takie jak rozdzielczosc czasowa wynikajac z doboru dlugosci, niedokladniosc wynikajaca z grubosc materialu czy przerwy maja drugorzedne znaczenie. Ma to odbicie zarowno w 
It can be concluded that the image quality increases for the higher two-gamma sensitivity scanners and is manifested mainly in the lower background variability values due to the higher statistics acquired. A similar effect was observed in analysing the data from the uExplorer TB PET scanner~\cite{zhang2017quantitative}.
The two-gamma sensitivity and XCAT image reconstruction analyses in terms of the quality metrics, together with the visual inspection of the reconstructed images show the advantage of longer, seven-ring S4 and S5 scanners. 
However, the improvement comes with a higher price. As estimated, the cost of the scintillator materials and SiPMs which constitute a majority of the TB price is more than two times higher for the S4, and S5 modalities compared to the three-ring solutions.
Still, the relative cost for all the scanners is about 10 to 4 times lower compared to the cost of the uExplorer modality. 
% At the same time S4, S5  are more costly, by % percent. 

%positronium
% tutaj moze odnosnik, ze to jest nowa metoda, ze trzeba dzialac w trybie single mode, (J-PET dziala w trigereless mode) wiec nawet jezeli dla innych skanerow jest dostepna to niekonieniczie. Podane wartosci sensitivity as wystarczajace dla pozytronium czas zycia.  
The importance of the high-sensitivity systems becomes more pronounced in the case of the positronium mean lifetime tomography which requires the registration of triple gamma coincidences.
The determined positronium sensitivity values make the J-PET TB modality a well-suited scanner for positronium lifetime imaging~\cite{moskalperspectives,moskal2024_ScienceSubmitted}. It is worth underlining that the positronium lifetime tomography is well suited to be performed on other high-sensitivity, large FOV scanners such as Biograph Vision Quadra or uExplorer. However, it requires operating on the single acquisition mode that permits to registration of multi-gamma coincidences. This condition can pose problems, especially for the system approved for clinical usage. 

The enumerated properties together with its cost-effectiveness and triggerless acquisition mode enabling three gamma positronium imaging, make the J-PET technology an attractive solution for broad application in clinics.

\section{Acknowledgments}

The authors gratefully acknowledge the support of the Foundation for Polish Science through programme TEAM POIR.04.04.00-00-4204/17; the National Science Centre of Poland through grant nos. 2019/35/B/ST2/03562, 2020/38/E/ST2/00112, 2021/42/A/ST2/00423 and 2021/43/B/ST2/02150; the Ministry of Education and Science under the grant No. SPUB/SP/490528/2021 
and IAL/SP/596235/2023; EU Horizon 2020 research and innovation programme, STRONG-2020 project, under grant agreement No 824093; the Jagiellonian University via the project CRP/0641.221.2020, and via SciMat and qLife Priority Research Areas under the program Excellence Initiative-Research University at the Jagiellonian University.

\section{Conflict of Interest Statement}

The Authors declare no conflict of interest.

\section*{References}
\addcontentsline{toc}{section}{\numberline{}References}
\vspace*{-20mm}

\listoffigures


\begin{thebibliography}{10}

\bibitem{schmall2019}
{Schmall~J.~P., Karp~J.~S. and Alavi~A.},
\newblock The potential role of total body PET imaging in assessment of atherosclerosis,
\newblock PET clinics {\bf 14}, 245--250 (2019).

\bibitem{alavi2021unparalleled}
A.~Alavi, T.~J. Werner, E.~St{\k{e}}pie{\'n}, and P.~Moskal,
\newblock Unparalleled and revolutionary impact of PET imaging on research and day to day practice of medicine,
\newblock Bio-Algorithms and Med-Systems {\bf 17}, 203--212 (2021).

\bibitem{alavi2020update}
A.~Alavi, S.~Hess, T.~J. Werner, and P.~F. H{\o}ilund-Carlsen,
\newblock An update on the unparalleled impact of FDG-PET imaging on the day-to-day practice of medicine with emphasis on management of infectious/inflammatory disorders, 2020.

\bibitem{Cherry2017}
{Cherry~S.~R., Badawi~R.~D., Karp~J.~S. et al.},
\newblock Total-body imaging: Transforming the role of positron emission tomography,
\newblock Science Translational Medicine {\bf 9}, eaaf6169 (2017).

\bibitem{spencer2021performance}
B.~A. Spencer et~al.,
\newblock Performance evaluation of the uEXPLORER total-body PET/CT scanner based on NEMA NU 2-2018 with additional tests to characterize PET scanners with a long axial field of view,
\newblock Journal of Nuclear Medicine {\bf 62}, 861--870 (2021).

\bibitem{karp2020}
{Karp~J.~S., Viswanath~V., Geagan~M.~J. et al.},
\newblock PennPET Explorer: design and preliminary performance of a whole-body imager,
\newblock Journal of Nuclear Medicine {\bf 61}, 136--143 (2020).

\bibitem{prenosil2022performance}
G.~A. Prenosil, H.~Sari, M.~F{\"u}rstner, A.~Afshar-Oromieh, K.~Shi, A.~Rominger, and M.~Hentschel,
\newblock Performance Characteristics of the Biograph Vision Quadra PET/CT System with a Long Axial Field of View Using the NEMA NU 2-2018 Standard,
\newblock Journal of nuclear medicine {\bf 63}, 476--484 (2022).

\bibitem{vandenberghe2020state}
S.~Vandenberghe, P.~Moskal, and J.~S. Karp,
\newblock State of the art in total body PET,
\newblock EJNMMI physics {\bf 7}, 1--33 (2020).

\bibitem{vandenberghe2022potential}
S.~Vandenberghe, N.~A. Karakatsanis, M.~A. Akl, J.~Maebe, S.~Surti, R.~A. Dierckx, D.~A. Pryma, S.~A. Nehmeh, O.~Bouhali, and J.~S. Karp,
\newblock The potential of a medium-cost long axial FOV PET system for nuclear medicine departments,
\newblock European Journal of Nuclear Medicine and Molecular Imaging {\bf 50}, 652--660 (2023).

\bibitem{zhang2020total}
X.~Zhang et~al.,
\newblock Total-body dynamic reconstruction and parametric imaging on the uEXPLORER,
\newblock Journal of Nuclear Medicine {\bf 61}, 285--291 (2020).

\bibitem{wang2022total}
G.~Wang, L.~Nardo, M.~Parikh, Y.~G. Abdelhafez, E.~Li, B.~A. Spencer, J.~Qi, T.~Jones, S.~R. Cherry, and R.~D. Badawi,
\newblock Total-Body PET Multiparametric Imaging of Cancer Using a Voxelwise Strategy of Compartmental Modeling,
\newblock Journal of Nuclear Medicine {\bf 63}, 1274--1281 (2022).

\bibitem{moskal2021positronium}
P.~Moskal et~al.,
\newblock Positronium imaging with the novel multiphoton PET scanner,
\newblock Science Advances {\bf 7}, eabh4394 (2021).

\bibitem{moskalperspectives}
P.~Moskal and E.~{\L}. St{\k{e}}pie{\'n},
\newblock Perspectives for translation of positronium imaging into clinics,
\newblock Frontiers in Physics , 891 (2022).

\bibitem{moskal2019feasibility}
{Moskal~P., Kisielewska~D., Curceanu~C. et al.},
\newblock {Feasibility study of the positronium imaging with the J-PET tomograph},
\newblock {Physics in Medicine and Biology} {\bf 64}, 055017 (2019).

\bibitem{beyene2023exploration}
E.~Y. Beyene et~al.,
\newblock Exploration of simultaneous dual-isotope imaging with multi-photon modular J-PET scanner,
\newblock Bio-Algorithms and Med-Systems {\bf 19} (2023).

\bibitem{Cherry2018}
{Cherry~S.~R., Jones~T., Karp~J.~S. et al.},
\newblock {Total-Body PET: Maximizing Sensitivity to Create New Opportunities for Clinical Research and Patient Care},
\newblock The Journal of Nuclear Medicine {\bf 59}, 3--12 (2018).

\bibitem{surti2020total}
{Surti~S., Pantel~A.~R. and Karp~J.~S.},
\newblock Total Body PET: Why, How, What for?,
\newblock IEEE Transactions on Radiation and Plasma Medical Sciences {\bf 4}, 283--292 (2020).

\bibitem{surti2013}
{Surti~S., Werner~M.~E. and Karp~J.~S.},
\newblock Study of PET scanner designs using clinical metrics to optimize the scanner axial FOV and crystal thickness,
\newblock Physics in Medicine and Biology {\bf 58}, 3995 (2013).

\bibitem{zhang2019sparse}
{Zhang~J., Knopp~M.~I. and Knopp~M.~V.},
\newblock Sparse detector configuration in SiPM digital photon counting PET: a feasibility study,
\newblock Molecular Imaging and Biology {\bf 21}, 447--453 (2019).

\bibitem{zein2020physical}
{Zein~S.~A., Karakatsanis~N.~A., Issa~M. et al.},
\newblock Physical performance of a long axial field-of-view PET scanner prototype with sparse rings configuration: A Monte Carlo simulation study,
\newblock Medical Physics {\bf 47}, 1949--1957 (2020).

\bibitem{cates2019}
{Cates~J.~W. and Levin~C.~S.},
\newblock Electronics method to advance the coincidence time resolution with bismuth germanate,
\newblock Physics in Medicine and Biology {\bf 64}, 175016 (2019).

\bibitem{gundacker2020experimental}
{Gundacker~S., Turtos~R.~M., Kratochwil~N., et al.},
\newblock Experimental time resolution limits of modern SiPMs and TOF-PET detectors exploring different scintillators and Cherenkov emission,
\newblock Physics in Medicine and Biology {\bf 65}, 025001 (2020).

\bibitem{Kowalski2018}
{Kowalski~P., Wi{\'s}licki~W., Shopa~R.~Y. et al.},
\newblock {Estimating the NEMA characteristics of the J-PET tomograph using the GATE package},
\newblock Physics in Medicine and Biology {\bf 63}, 165008 (2018).

\bibitem{moskal2021simulating}
P.~Moskal et~al.,
\newblock Simulating NEMA characteristics of the modular total-body J-PET scanner—an economic total-body PET from plastic scintillators,
\newblock Physics in Medicine and Biology {\bf 66}, 175015 (2021).

\bibitem{moskal2020prospects}
P.~Moskal and E.~St{\k{e}}pie{\'n},
\newblock Prospects and clinical perspectives of total-body PET imaging using plastic scintillators,
\newblock PET Clinics {\bf 15}, 439--452 (2020).

\bibitem{tayefi2023evaluation}
F.~Tayefi~Ardebili, S.~Nied{\'z}wiecki, and P.~Moskal,
\newblock Evaluation of modular J-PET sensitivity,
\newblock Bio-Algorithms and Med-Systems {\bf 19} (2023).

\bibitem{hiesmayr2019witnessing}
{Hiesmayr~B.~C. and Moskal~P.},
\newblock {Witnessing entanglement in Compton scattering processes via mutually unbiased bases},
\newblock Scientific Reports {\bf 9}, 1--14 (2019).

\bibitem{Moskal:2021kxe}
P.~Moskal et~al.,
\newblock {Testing CPT symmetry in ortho-positronium decays with positronium annihilation tomography},
\newblock Nature Commun. {\bf 12}, 5658 (2021).

\bibitem{moskal2024discrete}
P.~Moskal et~al.,
\newblock Discrete symmetries tested at 10- 4 precision using linear polarization of photons from positronium annihilations,
\newblock Nature Communications {\bf 15}, 78 (2024).

\bibitem{baran2019studies}
J.~Baran, J.~Gajewski, M.~Pawlik-Nied{\'z}wiecka, P.~Moskal, and A.~Ruci{\'n}ski,
\newblock Studies of J-PET detector to monitor range uncertainty in proton therapy,
\newblock in {\em 2019 IEEE Nuclear Science Symposium and Medical Imaging Conference (NSS/MIC)}, pages 1--4, IEEE, 2019.

\bibitem{baran2024feasibility}
J.~Baran et~al.,
\newblock Feasibility of the J-PET to monitor the range of therapeutic proton beams,
\newblock Physica Medica {\bf 118}, 103301 (2024).

\bibitem{borys2022protheramon}
D.~Borys et~al.,
\newblock ProTheRaMon—a GATE simulation framework for proton therapy range monitoring using PET imaging,
\newblock Physics in Medicine and Biology {\bf 67}, 224002 (2022).

\bibitem{brzezinski2023detection}
K.~W. Brzezinski et~al.,
\newblock Detection of range shifts in proton beam therapy using the J-PET scanner: a patient simulation study,
\newblock Physics in Medicine and Biology {\bf 68}, 145016 (2023).

\bibitem{shopa2021optimisation}
R.~Y. Shopa et~al.,
\newblock Optimisation of the event-based TOF filtered back-projection for online imaging in total-body J-PET,
\newblock Medical Image Analysis {\bf 73}, 102199 (2021).

\bibitem{raczynski20203d}
L.~Raczy{\'n}ski et~al.,
\newblock 3D TOF-PET image reconstruction using total variation regularization,
\newblock Physica Medica {\bf 80}, 230--242 (2020).

\bibitem{moskal2020performance}
{Moskal~P., Kisielewska~D., Shopa~R.~Y. et al.},
\newblock Performance assessment of the 2 $\gamma$ positronium imaging with the total-body PET scanners,
\newblock {EJNMMI Phys.} {\bf 7}, 44 (2020).

\bibitem{shibuya2020oxygen}
K.~Shibuya, H.~Saito, F.~Nishikido, M.~Takahashi, and T.~Yamaya,
\newblock Oxygen sensing ability of positronium atom for tumor hypoxia imaging,
\newblock Communications Physics {\bf 3}, 1--8 (2020).

\bibitem{yamaya2009multiplex}
T.~Yamaya, E.~Yoshida, N.~Inadama, F.~Nishikido, K.~Shibuya, M.~Higuchi, and H.~Murayama,
\newblock A multiplex “OpenPET” geometry to extend axial FOV without increasing the number of detectors,
\newblock IEEE Transactions on Nuclear Science {\bf 56}, 2644--2650 (2009).

\bibitem{daube2020performance}
M.~E. Daube-Witherspoon, V.~Viswanath, M.~E. Werner, and J.~S. Karp,
\newblock Performance characteristics of long axial field-of-view PET scanners with axial gaps,
\newblock IEEE Transactions on Radiation and Plasma Medical Sciences {\bf 5}, 322--330 (2020).

\bibitem{Smyrski2017}
J.~Smyrski et~al.,
\newblock Measurement of gamma quantum interaction point in plastic scintillator with WLS strips,
\newblock Nuclear Instruments and Methods in Physics Research Section A: Accelerators, Spectrometers, Detectors and Associated Equipment {\bf 851}, 39--42 (2017).

\bibitem{Niedzwiecki:2017nka}
S.~Niedźwiecki et~al.,
\newblock {J-PET: a new technology for the whole-body PET imaging},
\newblock Acta Phys. Polon. {\bf B48}, 1567 (2017).

\bibitem{kaplanoglu2023cross}
M.~T. Kaplanoglu and P.~Moskal,
\newblock A cross-staged gantry for total-body PET and CT imaging,
\newblock Bio-Algorithms and Med-Systems {\bf 19} (2023).

\bibitem{sarrut2021advanced}
D.~Sarrut et~al.,
\newblock Advanced Monte Carlo simulations of emission tomography imaging systems with GATE,
\newblock Physics in Medicine \& Biology {\bf 66}, 10TR03 (2021).

\bibitem{sarrut2022opengate}
D.~Sarrut et~al.,
\newblock The OpenGATE ecosystem for Monte Carlo simulation in medical physics,
\newblock Physics in Medicine and Biology {\bf 67}, 184001 (2022).

\bibitem{agostinelli2003geant4}
S.~Agostinelli et~al.,
\newblock GEANT4—a simulation toolkit,
\newblock Nuclear instruments and methods in physics research section A: Accelerators, Spectrometers, Detectors and Associated Equipment {\bf 506}, 250--303 (2003).

\bibitem{segars20104d}
W.~P. Segars, G.~Sturgeon, S.~Mendonca, J.~Grimes, and B.~M. Tsui,
\newblock 4D XCAT phantom for multimodality imaging research,
\newblock Medical physics {\bf 37}, 4902--4915 (2010).

\bibitem{zincirkeser2007standardized}
S.~Zincirkeser, E.~{\c{S}}ahin, M.~Halac, and S.~Sager,
\newblock Standardized uptake values of normal organs on 18F-fluorodeoxyglucose positron emission tomography and computed tomography imaging,
\newblock Journal of international medical research {\bf 35}, 231--236 (2007).

\bibitem{silva2015simulated}
J.~Silva-Rodr{\'\i}guez, P.~Aguiar, I.~Dom{\'\i}nguez-Prado, P.~Fierro, and {\'A}.~Ruibal,
\newblock Simulated FDG-PET studies for the assessment of SUV quantification methods,
\newblock Revista Espa{\~n}ola de Medicina Nuclear e Imagen Molecular (English Edition) {\bf 34}, 13--18 (2015).

\bibitem{raczynski2017calculation}
L.~Raczy{\'n}ski et~al.,
\newblock Calculation of the time resolution of the J-PET tomograph using kernel density estimation,
\newblock Physics in Medicine \& Biology {\bf 62}, 5076 (2017).

\bibitem{total_body_pet}
S.~Vandenberghe, P.~Moskal, and J.~Karp,
\newblock {State of the art in total body PET},
\newblock EJNMMI-Physics {\bf 7}, 35 (2020).

\bibitem{Thatcher558P}
J.~Thatcher and M.~Osman,
\newblock Prevalence, challenges and solutions for F-18 FDG PET/CT imaging of claustrophobic patients,
\newblock Journal of Nuclear Medicine {\bf 47}, 558P--558P (2006).

\bibitem{Moskal:2014sra}
P.~Moskal et~al.,
\newblock {Test of a single module of the J-PET scanner based on plastic scintillators},
\newblock Nucl. Instrum. Meth. {\bf A764}, 317--321 (2014).

\bibitem{moskal2016time}
P.~Moskal et~al.,
\newblock Time resolution of the plastic scintillator strips with matrix photomultiplier readout for J-PET tomograph,
\newblock Physics in Medicine \& Biology {\bf 61}, 2025 (2016).

\bibitem{moskal2016potential}
{Moskal~P., Alfs~D., Bednarski~T. et al.},
\newblock Potential of the {J-PET} detector for studies of discrete symmetries in decays of positronium atom-a purely leptonic system,
\newblock Acta Phys. Polon. {\bf B47}, 509 (2016).

\bibitem{moskal2018feasibility}
{Moskal~P., Krawczyk~N., Hiesmayr~B.~C. et al.},
\newblock Feasibility studies of the polarization of photons beyond the optical wavelength regime with the {J-PET} detector,
\newblock The European Physical Journal C {\bf 78}, 970 (2018).

\bibitem{zhang2017quantitative}
X.~Zhang, J.~Zhou, S.~R. Cherry, R.~D. Badawi, and J.~Qi,
\newblock Quantitative image reconstruction for total-body PET imaging using the 2-meter long EXPLORER scanner,
\newblock Physics in Medicine \& Biology {\bf 62}, 2465 (2017).

\bibitem{merlin2018castor}
T.~Merlin, S.~Stute, D.~Benoit, J.~Bert, T.~Carlier, C.~Comtat, M.~Filipovic, F.~Lamare, and D.~Visvikis,
\newblock CASToR: a generic data organization and processing code framework for multi-modal and multi-dimensional tomographic reconstruction,
\newblock Physics in Medicine \& Biology {\bf 63}, 185005 (2018).

\bibitem{shopa2023tof}
R.~Shopa et~al.,
\newblock TOF MLEM Adaptation for the Total-Body J-PET with a Realistic Analytical System Response Matrix,
\newblock IEEE Transactions on Radiation and Plasma Medical Sciences  (2023).

\bibitem{daube2006influence}
M.~E. Daube-Witherspoon, S.~Surti, S.~Matej, M.~Werner, S.~Jayanthi, and J.~S. Karp,
\newblock Influence of time-of-flight kernel accuracy in TOF-PET reconstruction,
\newblock in {\em 2006 IEEE Nuclear Science Symposium Conference Record}, volume~3, pages 1723--1727, IEEE, 2006.

\bibitem{NEMA:2018}
NEMA,
\newblock NEMA Standards Publication NU 2-2018: Performance Measurements of Positron Emission Tomographs (PET), 2018,
\newblock National Electrical Manufacturers Association (NEMA NU 2-2018).

\bibitem{efron1992bootstrap}
B.~Efron,
\newblock Bootstrap methods: another look at the jackknife,
\newblock in {\em Breakthroughs in statistics: Methodology and distribution}, pages 569--593, Springer, 1992.

\bibitem{efron_introduction_1993}
B.~Efron and R.~J. Tibshirani,
\newblock {\em An Introduction to the Bootstrap},
\newblock Springer {US}.

\bibitem{dahlbom2001estimation}
M.~Dahlbom,
\newblock Estimation of image noise in PET using the bootstrap method,
\newblock in {\em 2001 IEEE Nuclear Science Symposium Conference Record (Cat. No. 01CH37310)}, volume~4, pages 2075--2079, IEEE, 2001.

\bibitem{lartizien_comparison_2010}
C.~Lartizien, J.-B. Aubin, and I.~Buvat,
\newblock Comparison of Bootstrap Resampling Methods for 3-D {PET} Imaging,
\newblock  {\bf 29}, 1442--1454.

\bibitem{markiewicz2015assessment}
P.~Markiewicz, A.~Reader, and J.~Matthews,
\newblock Assessment of bootstrap resampling performance for PET data,
\newblock Physics in Medicine and Biology {\bf 60}, 279--299 (2015).

\bibitem{moskal2024_ScienceSubmitted}
P.~Moskal et~al.,
\newblock {First positronium image of the human brain in vivo},
\newblock medRxiv preprint medRxiv:https://doi.org/10.1101/2024.02.01.23299028  (2024).

\bibitem{kopka2020reconstruction}
P.~Kopka and K.~Klimaszewski,
\newblock RECONSTRUCTION OF THE NEMA IEC BODY PHANTOM FROM J-PET TOTAL-BODY SCANNER SIMULATION USING STIR.,
\newblock Acta Physica Polonica B {\bf 51}, 357 (2020).

\end{thebibliography}
\end{document}